\documentclass[twocolumn,amsmath,amssymb]{revtex4}

\usepackage{graphicx}
\usepackage{epstopdf}
\usepackage[colorlinks,citecolor=blue,linkcolor=blue]{hyperref}
\usepackage{tabularx}

\numberwithin{equation}{section}

\begin{document}

\title{\boldmath
Determination of $f_+^\pi(0)$ and Extraction of
$|V_{cd}|$ from Semileptonic $D$ Decays}

\author{G. Rong}
\author{Y. Fang}
\author{H. L. Ma}
\author{J. Y. Zhao}
\affiliation{Institute of High Energy Physics, Beijing 100039, People's Republic of China}

\date{\today}

\begin{abstract}
   By globally analyzing all existing measured branching fractions for $D\to \pi e^+\nu_e$ decays,
partial decay rates in different four momentum transfer-squared $q^2$ bins,
as well as products of the decay form factor $f_+^\pi(q^2)$ and the
Cabibbo-Kobayashi-Maskawa (CKM) quark-mixing matrix element $|V_{cd}|$,
we obtain $f_+^\pi(0)|V_{cd}|=0.1428\pm0.0019$.
This product, in conjunction with $|V_{cd}|$ from a global Standard Model fit,
implies a value for the $D\to\pi$ semileptonic form factor 
$f_+^\pi(0)=0.634\pm0.008\pm0.002$, 
which is consistent within error with those calculated in theory based on QCD,
but with higher precision than the most accurate
$f_+^\pi(0)_{\rm LQCD}=0.666\pm0.020\pm0.021$ calculated in LQCD by a factor of 3.3.
Alternately, using this product together with the most accurate form factor calculated in LQCD, 
we find $|V_{cd}|^{D\to \pi e^+\nu_e}=0.2144\pm0.0029_{\rm exp}\pm 0.0093_{\rm LQCD}$.
Combining this $|V_{cd}|^{D\to \pi e^+\nu_e}$
together with
$|V_{cd}|^{D^+\to\mu^+\nu_\mu}=0.2160\pm0.0049\pm0.0014$
extracted from both the BESIII and CLEO-c measurements
of $D^+\to\mu^+\nu_\mu$ decays,
we find the most precisely extracted $|V_{cd}|$ to be $|V_{cd}|=0.2157\pm0.0045$ up to date, 
which improves the accuracy
of the PDG'2014 value $|V_{cd}|_{\rm PDG'2014}=0.225\pm0.008$ by over $70\%$.
Using this $|V_{cd}|$ together with the PDG'2014 $|V_{ud}|$ and $|V_{td}|$,
we check for first column unitarity and find
$|V_{ud}|^2+|V_{cd}|^2+|V_{td}|^2-1=-0.004\pm0.002$,
which deviates from unitarity by $2\sigma$.
In addition, we find the ratio of $f_+^\pi(0)$ and $D^+$ decay constant $f_{D^+}$
to be $f_+^\pi(0)/f_{D^+}=(3.11\pm0.08)$ GeV$^{-1}$, 
which can be used to validate LQCD calculations for these two quantities.
\end{abstract}

\maketitle

\section{Introduction}
\label{sec:intro}

In the Standard Model (SM) of particle physics, the mixing between the quark flavors
in weak interaction is parameterized by the Cabibbo-Kobayashi-Maskawa (CKM) matrix $\hat V_{\rm CKM}$, 
which is a $3\times3$ unitary matrix.
Since the CKM matrix elements are fundamental parameters of the SM, 
they should be measured as accurately as possible.
Precise measurements of these elements are very important in testing the SM and searching for New Physics (NP) beyond the SM.
Any improved measurement of these elements would be the important input for precision test of the SM. 

Three generation unitarity can be checked to see whether
$\hat V_{\rm CKM} * \hat V_{\rm CKM}^\dagger = \hat I$ 
is satisfied,
which leads to test first, second and third column/row unitarity.
The unitarity also gives rise to unitarity triangle (UT) relation
$V_{ud}V_{ub}^{*}+V_{cd}V_{cb}^{*}+V_{td}V_{tb}^{*}=0$.
To check for this column/row unitarity and the UT
relation, many experimental measurements 
and theoretical efforts have been made in flavor physics 
for many years.
If any of these consistency checks significantly deviate from unitarity,
it may indicate some evidence for NP effects.

Each matrix element can be extracted from measurements of different processes 
supplemented by theoretical calculations for corresponding hadronic matrix elements.
Since the effects of strong interactions and weak interaction can be well separated in
semileptonic $D$ decays,
these decays are excellent processes from which one can determine the magnitude of the CKM matrix element $V_{cd(s)}$.
In the SM, neglecting the lepton mass, the differential decay rate for $D\to \pi e^+\nu_e$ process is given by
\begin{equation}\label{eq:dG_dq2}
    \frac{d\Gamma}{dq^2} = X \frac{G_F^2}{24\pi^3}|V_{cd}|^2 \boldsymbol{p}^3|f_+^\pi(q^2)|^2,
\end{equation}
where
$G_F$ is the Fermi constant, $\boldsymbol p$ is the three momentum of the $\pi$ meson
in the rest frame of the $D$ meson, $q^2$ is the four momentum transfer-squared,
i.e. the invariant mass of the lepton and neutrino system,
and $f_+^\pi(q^2)$ is the form factor which parameterizes the effect of strong interaction in the decay.
In Eq.~(\ref{eq:dG_dq2}), $X$ is a multiplicative factor due to isospin, which equals to 1 for mode $D^0\to\pi^-e^+\nu_e$
and  1/2 for mode $D^+\to\pi^0e^+\nu_e$.

In addition to extraction of $|V_{cd}|$,
precise measurements of the $D\to \pi$ semileptonic form factor is also very important to validate the lattice QCD (LQCD)
calculation of the form factor. If the LQCD calculation of the form factor pass the test with the precisely measured form factor
of $D\to \pi e^+\nu_e$ decay, the uncertainty of the semileptonic $B$ decay form factor calculated in LQCD would be reduced.
This would help in reducing the uncertainty of the measured $|V_{ub}|$ from semileptonic $B$ decays~\cite{RongG_program_ccbar}.
The improved measurement of $|V_{ub}|$ from semileptonic $B$ decay will improve the determination 
of the $B_d$ UT, 
from which one can more precisely test the SM and search for NP beyond the SM.

In the past decades, copious measurements of decay branching fractions and/or decay rates
for $D\to \pi e^+\nu_e$  were performed at different experiments.
To obtain the knowledge about $f_+^\pi(0)$ and
$|V_{cd}|$ as better as possible, we analyze all of these existing measurements.
By a comprehensive analysis of these existing measurements together with $|V_{cd}|$ from a global SM fit or
together with the form factor $f_+^\pi(0)$ calculated in LQCD,
we precisely determine $f_+^\pi(0)$ and extract $|V_{cd}|$.

In this article, we report the determination of $f_+^\pi(0)$ and extraction of $|V_{cd}|$
by analyzing all of these existing measurements of the
semileptonic $D\to \pi e^+\nu_e$ decays in conjunction with $|V_{cd}|$ from a global SM fit or with the form factor $f_+^\pi(0)$
calculated in LQCD.
In the following sections, we first review the experimental measurements of decay branching fractions and decay rates
for $D\to \pi e^+\nu_e$
and pre-deal with these measurements to get decay rates to be used in the comprehensive analysis
of all these existing measurements
in Section~\ref{sec:expt}.
We then describe our comprehensive analysis procedure for dealing with these measurements to obtain the product
of $f_+^\pi(0)$ and $|V_{cd}|$ in Section~\ref{sec:ana}.
In Section~\ref{sec:rslt}, we present the final results of our comprehensive analysis of these measurements.
We finally give a summary for the determination of $f_+^\pi(0)$ and the extraction of $|V_{cd}|$ in Section~\ref{sec:sum}.

\section{Experiments}
\label{sec:expt}

There are different kinds of measurements of $D\to \pi e^+\nu_e$ decays performed
at many experiments during last 25 years,
some of which can not directly be used to determine $f_+^\pi(0)$ and extract $|V_{cd}|$.
To determine these quantities from all of these existing measurements,
some of these measurements are needed to be pre-processed.

\subsection{Relative Measurements}

In 1995, by analyzing 3.0 fb$^{-1}$ data collected with the CLEO-II detector at the Cornell Electron Storage Ring (CESR),
the CLEO Collaboration made a measurement of the branching ratio of the Cabibbo suppressed $D^0$ semileptonic decays.
The CLEO Collaboration observed $87\pm33$ signal events for $D^0\to \pi^-e^+\nu_e$ decays and
obtained the ratio of branching fractions $R_0 \equiv B(D^0\to \pi^-e^+\nu_e)/B(D^0\to K^-e^+\nu_e)=0.103\pm0.039\pm0.013$~\cite{CLEO}.

The Cabibbo suppressed semileptonic $D^0\to \pi^-\ell^+\nu_\ell$ ($\ell=e,\mu$) decays
were studied at the E687 experiment in 1996.
The E687 Collaboration observed $45.4\pm13.3$ and $45.6\pm11.8$ signal events
for $D^0\to \pi^-e^+\nu_e$ and $D^0\to\pi^-\mu^+\nu_\mu$ decays, respectively.
After making a small correction to the muon events, the E687 Collaboration combined
the branching ratio measurements for the electron and muon modes together and determined the ratio of
decay branching fractions to be
$R_0 \equiv B(D^0\to \pi^-e^+\nu_e)/B(D^0\to K^-e^+\nu_e)=0.101\pm0.020\pm0.003$~
\cite{E687}.

By analyzing 4.8 fb$^{-1}$ data taken with the CLEO-II detector,
the CLEO Collaboration performed a measurement of the branching fraction for $D^+\to\pi^0e^+\nu_e$ decay.
The CLEO Collaboration found $65\pm15\pm20$ signal events for $D^+\to\pi^0e^+\nu_e$ decay and obtained the ratio
of the branching fractions to be
$R_+\equiv B(D^+\to\pi^0e^+\nu_e)/B(D^+\to\bar K^0e^+\nu_e)$
$=(4.5\pm1.6\pm1.9)\%$~\cite{CLEOII_Dp} in 1997.

In 2005, the CLEO Collaboration measured the branching ratios of the semileptonic $D^0\to\pi^-\ell^+\nu_\ell$
decays by analyzing about 7 fb$^{-1}$ of data collected
around the $\Upsilon(4S)$ resonance with the CLEO-III detector.
Combining their measurements for electron mode and muon mode with considering the differences
in phase spaces of these two decay modes, the CLEO Collaboration obtained
the ratio of branching fractions to be
$R_0 \equiv B(D^0\to \pi^-e^+\nu_e)$ $/B(D^0\to K^-e^+\nu_e)$ $=0.082\pm0.006\pm0.005$~\cite{CLEOIII}.

All above mentioned measurements are relative measurements which could not be used directly to determine
$f_+^\pi(0)$ or $|V_{cd}|$. To use these measurements to determine $f_+^\pi(0)$ or $|V_{cd}|$,
we should first transfer these measurements into absolute decay rates in certain $q^2$ range.
The absolute decay rate $\Delta\Gamma$ can be obtained from the measured relative decay branching ratio $R$ by
\begin{equation}
\Delta\Gamma=R\times B(D\to Ke^+\nu_e)\times\frac{1}{\tau_D},
\end{equation}
where
$B(D\to Ke^+\nu_e)$ is the branching fraction for $D^0\to K^-e^+\nu_e$ or $D^+\to \bar K^0e^+\nu_e$ decays,
and $\tau_D$ is the lifetime of $D$ meson.
Using the lifetime of $D$ meson, $\tau_{D^0}=(410.1\pm1.5)\times10^{-15}$ s, and $\tau_{D^+}=(1040\pm7)\times10^{-15}$ s,
the branching fractions of $B(D^0\to K^-e^+\nu_e)=(3.50\pm0.05)\%$ and $B(D^+\to \bar K^0e^+\nu_e)=(8.83\pm0.22)\%$
quoted from PDG'2014~\cite{pdg}, we translate these measurements of relative branching fractions into absolute partial decay rates
as shown in Tabs.~\ref{tab:Expt_D0} and \ref{tab:Expt_Dp}.

In 2014, the BaBar Collaboration studied the $D^0\to \pi^-e^+\nu_e$ decays by analyzing
347.2 fb$^{-1}$ data collected at 10.6 GeV~\cite{BaBar}.
They selected $D^0\to \pi^-e^+\nu_e$ decays
from $e^+e^-\to c\bar c$ events and divide the candidate events into ten $q^2$ bins.
In each $q^2$ bin, the branching fraction is measured
relative to the normalization mode, $D^0\to K^-\pi^+$.
The partial decay rate in $i$th $q^2$ bin is given by
\begin{equation}\label{eq:DeltaG_DeltaB_BaBar}
  \Delta\Gamma_i = \Delta B_i \times \frac{1}{\tau_{D^0}},
\end{equation}
where $\Delta B_i$ is the branching fraction measured in $i$th $q^2$ bin.
Inserting the lifetime of $D^0$ meson, $\tau_{D^0}=(410.1\pm1.5)\times10^{-15}$ s and the branching fraction values presented in Ref.~\cite{BaBar}
into Eq.~(\ref{eq:DeltaG_DeltaB_BaBar}),
we translate these measurements of branching fractions in ten $q^2$ bins into absolute partial decay rates, which are shown in Tab.~\ref{tab:Expt_D0}.

\begin{table}[h]
\newcolumntype{L}{>{\hsize=1\hsize\raggedright\arraybackslash}X}%
\newcolumntype{R}{>{\hsize=1\hsize\raggedleft\arraybackslash}X}%
\newcolumntype{C}{>{\hsize=1\hsize\centering\arraybackslash}X}%
\caption{The partial rates $\Delta\Gamma$ of the $D^0\to \pi^-e^+\nu_e$ decays in $q^2$ ranges obtained from different experiments.
$q^2_{\rm max}$ is the maximum value of $q^2$.
}
\label{tab:Expt_D0}
\begin{tabularx}{\linewidth}{XXC}
\hline\hline
Experiment & $q^2$ (GeV$/c^2$) & $\Delta\Gamma$ (ns$^{-1}$)  \\
\hline
 CLEO-II~\cite{CLEO}     & (0.0, $q^2_{\rm max}$) & $8.79\pm 3.51$ \\
 E687~\cite{E687}        & (0.0, $q^2_{\rm max}$) & $8.62\pm 1.73$ \\
 CLEO-III~\cite{CLEOIII} & (0.0, $q^2_{\rm max}$) & $7.00\pm 0.67$ \\
\hline
 BaBar~\cite{BaBar}
 & (0.0, 0.3) & $1.23\pm0.07$ \\
 & (0.3, 0.6) & $1.14\pm0.09$ \\
 & (0.6, 0.9) & $1.11\pm0.08$ \\
 & (0.9, 1.2) & $0.93\pm0.07$ \\
 & (1.2, 1.5) & $0.74\pm0.07$ \\
 & (1.5, 1.8) & $0.65\pm0.07$ \\
 & (1.8, 2.1) & $0.51\pm0.07$ \\
 & (2.1, 2.4) & $0.30\pm0.06$ \\
 & (2.4, 2.7) & $0.12\pm0.05$ \\
 & (2.7, $q^2_{\rm max}$) & $ 0.02\pm0.02$ \\
\hline
 Mark-III~\cite{MarkIII}      & (0.0, $q^2_{\rm max}$) & $9.51\pm4.26$ \\
 BES-II~\cite{BESII_D0}       & (0.0, $q^2_{\rm max}$) & $8.05\pm3.25$ \\
\hline
 CLEO-c~\cite{CLEOc}
 & (0.0, 0.3) & $1.39\pm0.10$ \\
 & (0.3, 0.6) & $1.22\pm0.09$ \\
 & (0.6, 0.9) & $1.02\pm0.08$ \\
 & (0.9, 1.2) & $0.98\pm0.08$ \\
 & (1.2, 1.5) & $0.79\pm0.07$ \\
 & (1.5, 2.0) & $0.84\pm0.07$ \\
 & (2.0, $q^2_{\rm max}$) & $0.80\pm0.07$ \\
\hline\hline
\end{tabularx}
\end{table}

\begin{table}[h]
\newcolumntype{L}{>{\hsize=1\hsize\raggedright\arraybackslash}X}%
\newcolumntype{R}{>{\hsize=1\hsize\raggedleft\arraybackslash}X}%
\newcolumntype{C}{>{\hsize=1\hsize\centering\arraybackslash}X}%
\caption{The partial rates of the $D^+\to \pi^0 e^+\nu_e$ decays in $q^2$ ranges obtained from different experiments.
$q^2_{\rm max}$ is the maximum value of $q^2$.
}
\label{tab:Expt_Dp}
\begin{tabularx}{\linewidth}{XXC}
\hline\hline
Experiment  & $q^2$ (GeV$/c^2$) & $\Delta\Gamma$ (ns$^{-1}$) \\
\hline
 CLEO-II~\cite{CLEOII_Dp}  & (0.0, $q^2_{\max}$) & $3.82\pm2.11$ \\
\hline
 CLEO-c~\cite{CLEOc}
  & (0.0, 0.3) & $0.71\pm0.07$ \\
  & (0.3, 0.6) & $0.66\pm0.07$ \\
  & (0.6, 0.9) & $0.56\pm0.07$ \\
  & (0.9, 1.2) & $0.57\pm0.07$ \\
  & (1.2, 1.5) & $0.48\pm0.07$ \\
  & (1.5, 2.0) & $0.54\pm0.07$ \\
  & (2.0, $q^2_{\rm max}$) & $0.37\pm0.07$ \\
\hline\hline
\end{tabularx}
\end{table}

\subsection{Absolute Measurements}

In 1989, the Mark III Collaboration performed a measurement of absolute branching fraction for semileptonic $D^0\to \pi^-e^+\nu_e$ decay by analyzing data taken at the peak of $\psi(3770)$ resonance with the Mark III detector.
They tagged $3636\pm54\pm195$ $\bar D^0$ mesons and found 7 $D^0\to \pi^-e^+\nu_e$ signal events in the system recoiling against the $\bar D^0$ tags.
With these events, they measured the absolute decay branching fraction $B(D^0\to \pi^-e^+\nu_e)=(0.39^{+0.23}_{-0.11}\pm0.04)\%$~\cite{MarkIII}.

Using the similar method as the one used in Mark III, the BES-II Collaboration measured the branching fractions
of $D^0\to \pi^-e^+\nu_e$ decays by analyzing about 33 pb$^{-1}$ data taken around 3.773 GeV
with the BES-II detector at the BEPC collider.
In the system recoiling against the $\bar D^0$ tags, $9.0\pm3.6$ events from $D^0\to\pi^-e^+\nu_e$ decays were observed.
With these events, the branching fraction is measured to be
$B(D^0\to\pi^-e^+\nu_e)=(0.33\pm0.13\pm0.03)\%$~\cite{BESII_D0}.

The partial decay rate relates to the decay branching fraction by
\begin{equation}
\Delta\Gamma=B(D^0\to \pi^- e^+\nu_e)\times\frac{1}{\tau_{D^0}}.
\end{equation}
Using the lifetime of $D^0$ meson quoted from PDG'2014~\cite{pdg},
$\tau_{D^0}=(410.1\pm1.5)\times10^{-15}$ s, we translate these absolute measurements of branching fractions
for $D^0\to \pi^-e^+\nu_e$ decays into the partial decay rates, which are shown in Tab.~\ref{tab:Expt_D0}.

In 2009, the CLEO Collaboration studied the semileptonic decays of $D^0\to \pi^-e^+\nu_e$ and $D^+\to \pi^0e^+\nu_e$ decays by analyzing 818 pb$^{-1}$ data
collected at 3.773 GeV with the CLEO-c detector. Using double tag method, they measured the decay rates for semileptonic $D^0\to \pi^-e^+\nu_e$ and $D^+\to \pi^0e^+\nu_e$ decays in seven $q^2$ bins~\cite{CLEOc}.
These measurements of decay rates are summarized in Tabs.~\ref{tab:Expt_D0} and \ref{tab:Expt_Dp}.

In 2006, the Belle Collaboration published results on the $D^0\to \pi^-\ell^+\nu_\ell$ decays.
They accumulated $56461\pm309\pm830$ inclusive $D^0$ mesons and found $126\pm12\pm3$ signal events
for $D^0\to \pi^-e^+\nu_e$ decays and $106\pm12\pm6$ signal events for $D^0\to \pi^-\mu^+\nu_\mu$ decays
from 282 fb$^{-1}$ data collected around 10.58 GeV with the Belle detector~\cite{Belle}.
Using these selected events from semileptonic $D^0$ decays, the Belle Collaboration obtained the
form factors $f_+^\pi(q^2)$ in ten $q^2$ bins with the bin size of 0.3 GeV$^2/c^4$.
To obtain the product $f_+^\pi(q^2_i)|V_{cd}|$ which will be used in our comprehensive analysis in Section~\ref{sec:ana},
we extrapolate these measurements of form factors at the Belle experiment to the product $f_+^\pi(q^2_i)|V_{cd}|$
using the PDG'2006 value of $|V_{cd}|=0.2271\pm0.0010$~\cite{pdg2006} which was originally used in the Belle's paper published.
Table~\ref{tab:ffVcd_Belle} lists the form factors $f_+^\pi(q^2_i)$ measured at the Belle experiment
and our translated products $f_+^\pi(q^2_i)|V_{cd}|$.
These products will be used in our further analysis described in Section~\ref{sec:ana}.

\begin{table}[!hbp]
\newcolumntype{L}{>{\hsize=1\hsize\raggedright\arraybackslash}X}%
\newcolumntype{R}{>{\hsize=1\hsize\raggedleft\arraybackslash}X}%
\newcolumntype{C}{>{\hsize=1\hsize\centering\arraybackslash}X}%
\caption{Measurements of form factors $f_+^\pi(q^2_i)$ at the Belle experiment and the products $f_+^\pi(q^2_i)|V_{cd}|$
obtained from the Belle and BESIII experiments.}
\label{tab:ffVcd_Belle}
\begin{tabularx}{\linewidth}{llCC}
\hline\hline
Experiment & $q^2_i$ (GeV$/c^2$) & $f_+^\pi(q^2_i)$ & $f_+^\pi(q^2_i)|V_{cd}|$ \\
\hline
Belle~\cite{Belle}
   &  0.15 & $0.637\pm0.053$ & $0.145\pm0.012$ \\
   &  0.45 & $0.797\pm0.067$ & $0.181\pm0.015$ \\
   &  0.75 & $0.853\pm0.077$ & $0.194\pm0.017$ \\
   &  1.05 & $0.830\pm0.090$ & $0.188\pm0.020$ \\
   &  1.35 & $0.963\pm0.107$ & $0.219\pm0.024$ \\
   &  1.65 & $0.940\pm0.143$ & $0.213\pm0.033$ \\
   &  1.95 & $1.430\pm0.190$ & $0.325\pm0.043$ \\
   &  2.25 & $1.760\pm0.273$ & $0.400\pm0.062$ \\
   &  2.55 & $1.820\pm0.447$ & $0.413\pm0.101$ \\
   &  2.85 & $2.157\pm1.243$ & $0.490\pm0.282$ \\
\hline
BESIII~\cite{BESIII_D0Kenu} & 0.0 & & $0.1420\pm0.0026$ \\
\hline\hline
\end{tabularx}
\end{table}

Recently, the BESIII Collaboration reported preliminary results of
$D^0\to \pi^-e^+\nu_e$ decays obtained by analyzing
2.92 fb$^{-1}$ data taken at 3.773 GeV.
The BESIII Collaboration accumulated $(279.3\pm0.4)\times10^4$ $\bar D^0$ tags
from five hadronic decay modes.
In this sample of $\bar D^0$ tags, they observed $6297\pm87$ signal events
for $D^0\to \pi^-e^+\nu_e$ decays
~\cite{BESIII_D0Kenu,CKM2014_RongG_Talk}, and
measured differential rates of $D^0\to \pi^-e^+\nu_e$ decays at nine
$q^2$ bins from 0.0 to 3.0 GeV$^2$.
By analyzing these differential decay rates
the BESIII collaboration measured
a value of the product~\cite{BESIII_D0Kenu,CKM2014_RongG_Talk}
$$f_+^\pi(0)|V_{cd}|=0.1420\pm 0.0024\pm 0.0010,$$
which is obtained from a fit to the data in the case of
that the form of form factor is parameterized with three-parameters series expansion (see Section~\ref{sec:ana}).
The last row of Tab.~\ref{tab:ffVcd_Belle} lists this $f_+^\pi(0)|V_{cd}|$, where the error is the combined
statistical and systematic errors.


\section{Analysis}
\label{sec:ana}

To obtain the product of the semileptonic form factor at four momentum transfer $q=0$,
$f^\pi_+(0)$, and the magnitude of CKM matrix element $|V_{cd}|$,
we perform a comprehensive $\chi^2$ fit to
these experimental measurements of the partial decay rates 
and the products $f_+^\pi(q^2_i)|V_{cd}|$ listed in Tabs.~\ref{tab:Expt_D0},~\ref{tab:Expt_Dp} and~\ref{tab:ffVcd_Belle}.
The object function to be minimized in the fit is defined as
\begin{equation}\label{eq:chi2}
    \chi^2 = \chi^2_{\rm R} + \chi^2_{\rm P},
\end{equation}
where
$\chi^2_{\rm R}$ is
for both the decay rates extrapolated from measurements of decay branching fraction
and the partial decay rates measured in different $q^2$ ranges,
and
$\chi^2_{\rm P}$ corresponds to the products of $f_+^\pi(q^2_i)|V_{cd}|$
obtained from Belle's measurements of $f_+^\pi(q^2_i)$ and $f_+^\pi(0)|V_{cd}|$ measured at the BESIII experiment.

Taking into account the correlations between the measurements of the partial decay rates, the quantity $\chi^2_{\rm R}$ is given by
\begin{equation}\label{eq:chi2_r}
    \chi^2_{\rm R} = \sum_{i=1}^{30}\sum_{j=1}^{30} (\Delta\Gamma^{\rm ex}_i - \Delta\Gamma^{\rm th}_i) (\mathcal C^{-1}_{\rm R})_{ij} (\Delta\Gamma^{\rm ex}_j - \Delta\Gamma^{\rm th}_j),
\end{equation}
where
$\Delta\Gamma^{\rm ex}$ denotes the experimentally measured partial decay rate, $\Delta\Gamma^{\rm th}$
is the theoretical expectation of the decay rate, and
$\mathcal C^{-1}_{\rm R}$ is the inverse of the covariance matrix
$\mathcal C_{\rm R}$, which is a $30\times30$ matrix containing the correlations
between the measured partial decay rates listed in Tabs.~\ref{tab:Expt_D0} and ~\ref{tab:Expt_Dp}.
The construction of $C_{\rm R}$ is discussed in subsection~\ref{sec:cov}.
With the parametrization of the form factor, the theoretically predicted partial decay rate
in a given $q^2$ bin is obtained by integrating Eq.~(\ref{eq:dG_dq2})
from the low boundary $q^2_{\rm low}$ to the up boundary $q^2_{\rm up}$ of the $q^2$ bin,
\begin{equation}\label{eq:DR_th}
    \Delta\Gamma^{\rm th} = \int_{q^2_{\rm low}}^{q^2_{\rm up}} X \frac{G_F^2}{24\pi^3}|V_{cd}|^2 \boldsymbol{p}^3|f_+^\pi(q^2)|^2 dq^2.
\end{equation}
In this analysis, we used several forms of the form factor parameterizations which are discussed in subsection~\ref{sec:form_facor}.

Ignoring some possible correlations of the measurements of the product $f_+^\pi(q^2_i)|V_{cd}|$ measured at the Belle and BESIII experiments,
the function $\chi^2_{\rm P}$ in Eq.~(\ref{eq:chi2}) is defined as
\begin{equation}\label{eq:chi2_P}
    \chi^2_{\rm P} = \sum_{i=1}^{11} \left( \frac{\tilde f_i^{\rm ex}-\tilde f_i^{\rm th}}{\sigma_i} \right)^2,
\end{equation}
where $\tilde f_i^{\rm ex}$ is the measured product $f_+^\pi(q^2)|V_{cd}|$ at $q^2_i$ with the standard deviation $\sigma_i$, and $\tilde f_i^{\rm th}$
is the theoretical expectation of the product $f_+^\pi(q^2)|V_{cd}|$ at $q^2_i$.

\subsection{Form Factor Parameterizations}
\label{sec:form_facor}

Several model dependent calculations of form factor are often used in analysis of experimental measurements of
semileptonic $D$ decays.

In general, the single pole model is the simplest approach to describe the $q^2$ dependent behavior of form factor.
The single pole model is expressed as
\begin{equation}\label{eq:ff_pole}
    f_+^\pi(q^2) = \frac{f_+^\pi(0)}{1-q^2/m_{\rm pole}^2},
\end{equation}
where $f_+^\pi(0)$ is the value of form factor at $q^2=0$, $m_{\rm pole}$ is the pole
mass which is predicted to be the mass of the $D^{*+}$ meson for semileptonic $D\to \pi \ell^+\nu_\ell$ decays.

The so-called BK parameterization~\cite{BK} is also widely used in lattice QCD calculations
and experimental studies of this decay. In the BK parameterization, the form factor
of the semileptonic $D\to \pi \ell^+\nu_\ell$ decays is written as
\begin{equation}\label{eq:ff_BK}
    f_+^\pi(q^2) = \frac{f_+^\pi(0)}{(1-q^2/m_{D^{*+}}^2)(1-\alpha q^2/m_{D^{*+}}^2)},
\end{equation}
where $m_{D^{*+}}$ is the mass of the $D^{*+}$ meson,
and $\alpha$ is a free parameter to be fitted.
The value of $\alpha$ is assumed to be around $1.34$
for $D\to \pi\ell^+\nu_\ell$ in the BK parameterization.

The ISGW2 model~\cite{ISGW2} assumes
\begin{equation}\label{eq:ff_ISGW2}
    f_+^\pi(q^2) = f_+^\pi(q^2_{\rm max}) \left( 1+\frac{r^2}{12}(q^2_{\rm max} - q^2) \right)^{-2},
\end{equation}
where $q^2_{\rm max}$ is the kinematical limit of $q^2$,
and $r$ is the conventional radius of the meson.
In this model, the prediction of $r$ for $D\to \pi\ell^+\nu_\ell$ decays is $1.410$ GeV$^{-1}c$.

The most general parameterization of the form factor is the series expansion~\cite{ff_zexpansion},
which is based on analyticity and unitarity.
In this parametrization, the variable $q^2$ is mapped to a new variable $z$ through
\begin{equation}
   z(q^2,t_0) = \frac{\sqrt{t_+-q^2}-\sqrt{t_+-t_0}}{\sqrt{t_+-q^2}+\sqrt{t_+-t_0}},
\end{equation}
with $t_{\pm}=(m_D\pm m_\pi)^2$ and $t_0 = t_+(1-\sqrt{1-t_-/t_+})$.
The form factor is then expressed in terms of the new variable $z$ as
\begin{equation}\label{eq:ff_series}
   f_+^\pi(q^2) = \frac{1}{P(q^2)\phi(q^2,t_0)} \sum_{k=0}^{\infty} a_k(t_0)[z(q^2,t_0)]^k,
\end{equation}
where $P(q^2)=1$ for $D\to\pi\ell^+\nu_\ell$,
$\phi(q^2,t_0)$ is an arbitrary function,
and $a_k(t_0)$ are real coefficients.
In this analysis, the choice of $\phi(q^2,t_0)$ is taken to be
\begin{eqnarray}
  \phi(q^2,t_0) &=& \left( \frac{\pi m^2_c}{3} \right)^{\frac12} \left( \frac{z(q^2,0)}{-q^2} \right)^{\frac52} \left( \frac{z(q^2,t_0)}{t_0-q^2} \right)^{-\frac12}
  \nonumber
  \\
  &\times& \left( \frac{z(q^2,t_-)}{t_--q^2} \right)^{-\frac34} \frac{(t_+-q^2)}{(t_+-t_0)^{\frac14}},
\end{eqnarray}
where $m_c$ is the mass of charm quark, which is taken to be $1.2$ GeV$/c^2$.
In practical use, one usually make a truncation on the above series.
Actually, it is found that the current experimental data can be adequately described by only the first three
terms in Eq.~(\ref{eq:ff_series}).

In this analysis we will fit the measured decay rates to the three-parameter series expansion.
After optimizing the form factor parameters, we obtain the form for the three-parameter series expansion:
\begin{equation}\label{eq:ff_3series}
    f_+^\pi(q^2) =  \frac{f_+^\pi(0)P(0)\phi(0,t_0) (1+\sum_{k=1}^{2}r_k [z(q^2,t_0)]^k)}{P(q^2)\phi(q^2,t_0) (1+\sum_{k=1}^{2}r_k [z(0,t_0)]^k)},
\end{equation}
where $r_k\equiv a_k(t_0)/a_0(t_0)$ ($k=1,2$).

\subsection{Covariance Matrix}
\label{sec:cov}

It's a little complicated to compute the covariances of these 30 measurements of partial decay rates in different $q^2$ ranges and at different experiments. To be clear, we separate the correlations among these $\Delta\Gamma$ measurements into two case: the one associated with the experimental status of each independent experiment, and the one related to the external inputs of parameters such as the lifetime of the $D$ meson.

The statistical uncertainties in the $\Delta\Gamma$ measurements from the same experiment are correlated to some extent, while these are independent for the measurements from different experiments.
The systematic uncertainties from tracking, particle identification, etc. are usually independent between different experiments.
In this analysis, we treat the systematic uncertainties except the ones from $D$ lifetimes and branching fractions for $D\to K e^+\nu_e$ as fully uncorrelated between the measurements performed at different experiments.
We consider these below:

\begin{itemize}
\item
The covariances of the $\Delta\Gamma$ measured at the same experiment are computed using the statistical errors,  the systematic errors, and the correlation coefficients, which are presented in their original papers published.

\item
For the measurements of $D^0\to \pi^-e^+\nu_e$ decay, the lifetime of $D^0$ meson is used to obtain the partial decay rates
in particular $q^2$ ranges. The systematic uncertainties due to imperfect knowledge of $D^0$ lifetime are fully correlated
among all these measurements of the partial rates of $D^0\to \pi^-e^+\nu_e$ decay.
Similarly, the systematic uncertainties related to $D^+$ lifetime are fully correlated
among all of the $\Delta\Gamma$ measurements for $D^+\to\pi^0e^+\nu_e$ decay.

\item
An additional systematic uncertainty from the branching fraction for $D^0\to K^-e^+\nu_e$ decay
is fully correlated between these relative measurements of $D^0\to \pi^-e^+\nu_e$ decay at the CLEO-II, E687 and CLEO-III experiments.
Since we only use one relative measurement of $D^+\to \pi^0e^+\nu_e$ decay which is from the CLEO-II experiment,
there are no correlations due to the branching fraction for $D^+\to\bar K^0e^+\nu_e$ between this measurement and other measurements.
\end{itemize}

With these considerations mentioned above, we then construct a $30\times30$ covariance matrix $\mathcal C_{\rm R}$ which is necessary in  the form factor fit.

\subsection{Fits to Experimental Data}

Four fits are applied to the experimental data with the form factor hypothesis of single pole model, BK model,
ISGW2 model and series expansion. The fit to experimental data returns the normalization $f_+^\pi(0)|V_{cd}|$
and the shape parameters of the form factor which govern the behavior of form factor in high $q^2$ range.

\begin{figure*}
  \includegraphics[width=\textwidth]{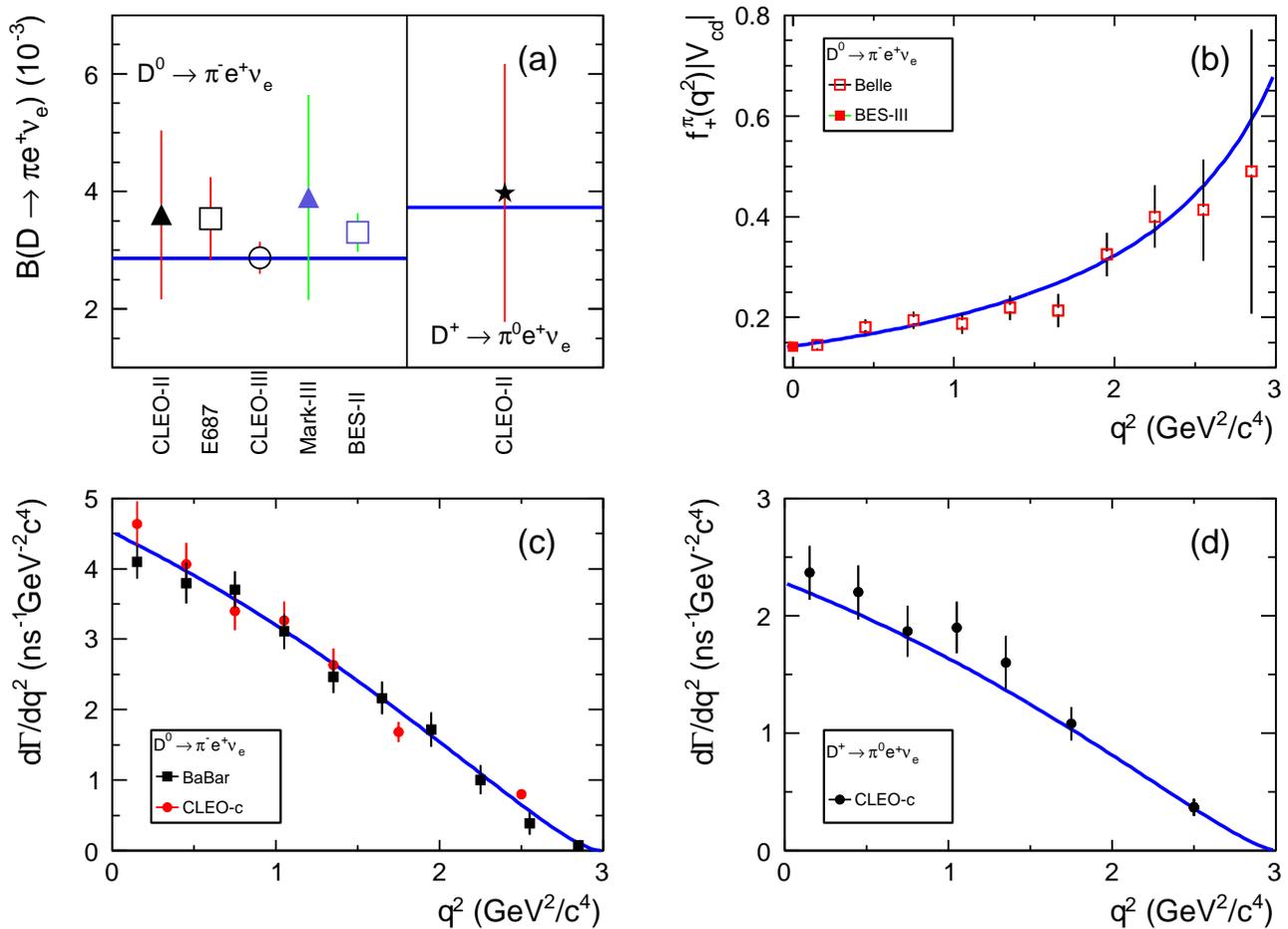}
  \caption{
  (a) Comparisons of branching fraction measurements for $D\to \pi e^+\nu_e$ decays,
  (b) the product $f_+^\pi(q^2)|V_{cd}|$ measured at the Belle and BESIII experiments,
  the differential decay rates as function of $q^2$ for (c) $D^0\to \pi^-e^+\nu_e$ measured at the BaBar and CLEO-c experiments,
      and (d) for $D^+\to \pi^0e^+\nu_e$ measured at the CLEO-c experiment.
  The blue lines show the fit to these measurements using the series expansion for the form factor.
  }
  \label{fig:fit}
\end{figure*}

The numerical results of the fit corresponding to each form of the form factor parameterization are summarized
in Tab.~\ref{tab:results}.
As an example, Fig.~\ref{fig:fit}
presents the result of the fit in the case of using the form factor parameterization of series expansion.
In Fig.~\ref{fig:fit} (a), we compared the measured branching fractions of $D\to \pi e^+\nu_e$ decays from different experiments.
Figure~\ref{fig:fit} (b) depicts the measurements of $f_+^\pi(q^2)|V_{cd}|$ at different $q^2$ from the Belle and BESIII experiments.
Figure~\ref{fig:fit} (c) and (d) show the measured differential decay rates for
$D^0\to \pi^-e^+\nu_e$ and $D^+\to \pi^0e^+\nu_e$, respectively.
In these figures, the lines show the best fit to these measurements of
$D\to \pi e^+\nu_e$ decays.

\begin{table*}
\newcolumntype{L}{>{\hsize=1\hsize\raggedright\arraybackslash}X}%
\newcolumntype{R}{>{\hsize=1\hsize\raggedleft\arraybackslash}X}%
\newcolumntype{C}{>{\hsize=1\hsize\centering\arraybackslash}X}%
  \centering
  \caption{Fitted parameters corresponding to different form factor parameterizations and $\chi^2/{\rm d.o.f.}$ of the fit.}
  \label{tab:results}
  \begin{tabularx}{\linewidth}{lCrLc}
  \hline\hline
  Parameterization & $f_+^\pi(0)|V_{cd}|$ & & Shape Parameters &  $\chi^2/{\rm d.o.f.}$ \\
  \hline
  Single pole      & $0.1447 \pm 0.0015$ & $M_{\rm pole}$ & $=(1.905 \pm 0.016)$ GeV$/c^2$ & $27.3/39$ \\
  BK               & $0.1429 \pm 0.0017$ & $\alpha$ & $= 0.252 \pm 0.044$ &  $25.5/39$ \\
  ISGW2            & $0.1417 \pm 0.0016$ & $r$ & $=(2.01 \pm 0.05)$ GeV$^{-1}c^2$ & $28.6/39$ \\
  Series expansion & $0.1428 \pm 0.0019$ & $r_1$ & $=-1.95 \pm 0.33$  & $25.0/38$ \\
  & & $r_2$ & $=-0.11 \pm 1.84$ &  \\
  \hline\hline
  \end{tabularx}
\end{table*}

\begin{figure*}
  \includegraphics[width=\textwidth]{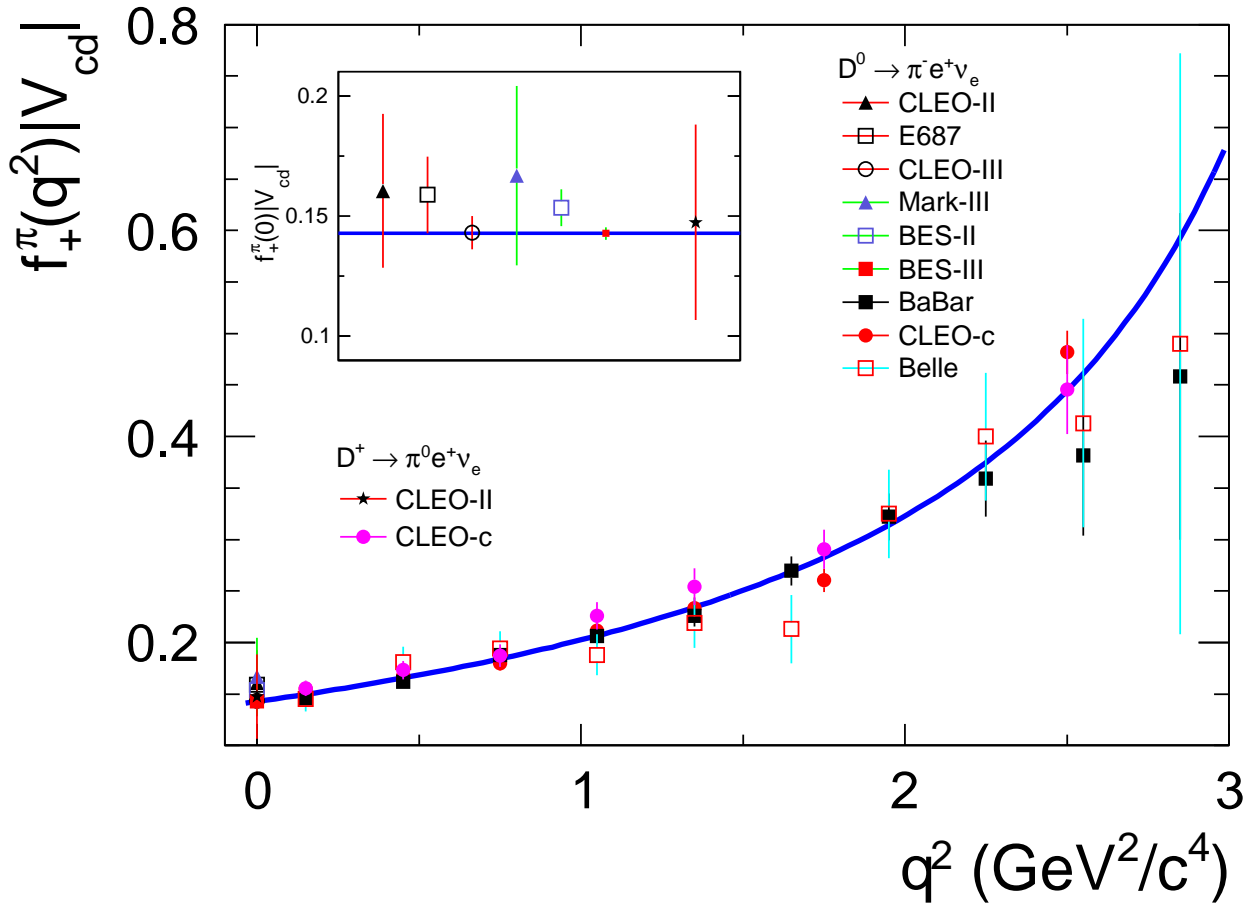}
  \caption{The product $f_+^\pi(q^2)|V_{cd}|$ measured at different experiments as a function of $q^2$. The blue curve represents the series expansion fit to these $f_+^\pi(q^2)|V_{cd}|$. The insert plot shows the comparison of the products $f_+^\pi(0)|V_{cd}|$ which are obtained using the branching fractions measured at different experiments. }
  \label{fig:ff_x_Vcd}
\end{figure*}

To check the fit quality and also the isospin invariance, the experimentally
measured decay branching fractions and/or partial rates are mapped into the product $f_+^\pi(q^2_i)|V_{cd}|$ via
\begin{equation}\label{eq:f0_B}
    f_+^\pi(0)|V_{cd}| = \sqrt{\frac{B}{\tau_D}\frac{1}{XN}}
\end{equation}
and
\begin{equation}\label{eq:f0_DR}
    f_+^\pi(q^2_i)|V_{cd}| = \sqrt{\left(\frac{d\Gamma}{dq^2}\right)_i\frac{24\pi^3}{XG_F^2\boldsymbol p_i^3}},
\end{equation}
where
$B$ denotes the measured branching fraction, the differential decay rate $(d\Gamma/dq^2)_i$ is obtained by dividing measured decay rate in $q^2$ bin $i$ by the corresponding bin size.
The normalization $N$ is given by
\begin{equation}\label{eq:norm}
    N = \frac{G_F^2}{24\pi^3|f_+^\pi(0)|^2}\int_{0}^{q^2_{\rm max}} {\boldsymbol p}^3|f_+^\pi(q^2)|^2dq^2.
\end{equation}
The effective $\boldsymbol p_i^3$ in $q^2$ bin $i$ is given by
\begin{equation}\label{eq:p3}
    \boldsymbol p_i^3 = \frac{\int_{q^2_{\rm low}}^{q^2_{\rm up}} {\boldsymbol p}^3|f_+^\pi(q^2)|^2dq^2}{|f_+^\pi(q^2_i)|^2(q^2_{\rm up}-q^2_{\rm low})}.
\end{equation}
To calculate the integral in Eqs.~(\ref{eq:norm}) and (\ref{eq:p3}), we use the shape parameters of the form factor,
which is obtained from the series expansion fit to the data.

Figure~\ref{fig:ff_x_Vcd} shows the product $f_+^\pi(q^2)|V_{cd}|$ as a function of $q^2$,
where
the blue curve corresponds to the best series expansion fit to the experimental data.
In this fit, seven measurements of $f_+^\pi(0)|V_{cd}|$ locate at $q^2=0$, which overlap each other.
To be clear, these $f_+^\pi(0)|V_{cd}|$ translated from the decay branching fractions measured at
different experiments are also displayed in the insert plot in Fig.~\ref{fig:ff_x_Vcd}.


\section{Results}
\label{sec:rslt}

In this analysis, we choose the results from the fit using series expansion as our primary results
and use this extracted $f_+^\pi(0)|V_{cd}|$ from the fit to determine the form factor $f_+^\pi(0)$
and extract the magnitude of the CKM matrix element $V_{cd}$.

\subsection{Form Factor $f_+^\pi(0)$}

\begin{figure}[h]
  \includegraphics[width=0.5\textwidth]{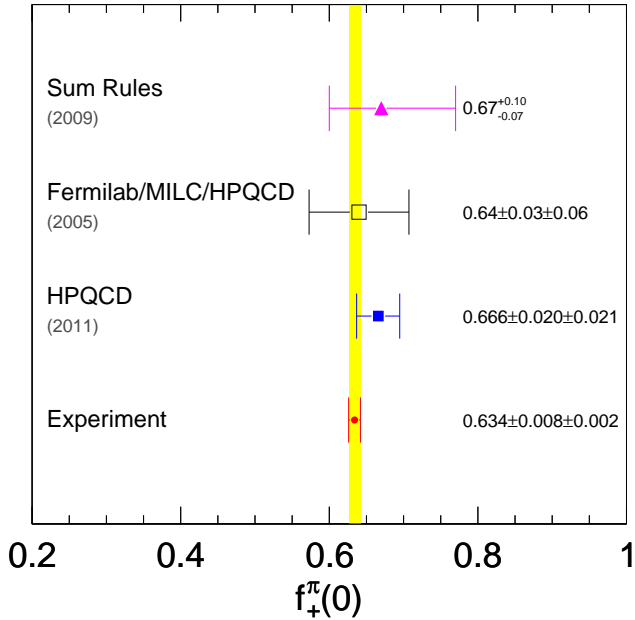}
  \caption{Comparison of our determined form factor from experimental measurements with the theoretical calculations of the form factor. }
  \label{fig:Cmp_f0}
\end{figure}

Dividing the value of
$$f_+^\pi(0)|V_{cd}|=0.1428\pm0.0019$$
shown in Tab.~\ref{tab:results} from the series expansion fit by the
$|V_{cd}|=0.22522\pm0.00061$ obtained from the global SM fit~\cite{pdg}
yields the form factor
\begin{equation}\label{eq:f0}
   f_+^\pi(0) = 0.634\pm0.008\pm0.002,
\end{equation}
where the first uncertainty is from the combined statistical and systematic uncertainties in the partial decay rate measurements,
and the second is due to the uncertainty in the $|V_{cd}|$.
The result of the form factor determined in this analysis is compared
with the theoretical calculations of the form factor from the lattice QCD ~\cite{HPQCD, Fermilab2005}
and from QCD light-cone sum rules~\cite{SR} in Fig.~\ref{fig:Cmp_f0}.
Our result of the form factor determined by analyzing all existing experimental measurements of these decays
is consistent within error with these values predicted by theory,
but is with higher precision than the most accurate value of the form factor,
$f_+^{\pi}(0)_{\rm LQCD}=0.666\pm0.020\pm0.021$ calculated in LQCD~\cite{HPQCD},
calculated in LQCD by a factor of 3.3.

\subsection{Parameters of Form Factor}

\begin{figure}[h]
  \includegraphics[width=0.5\textwidth]{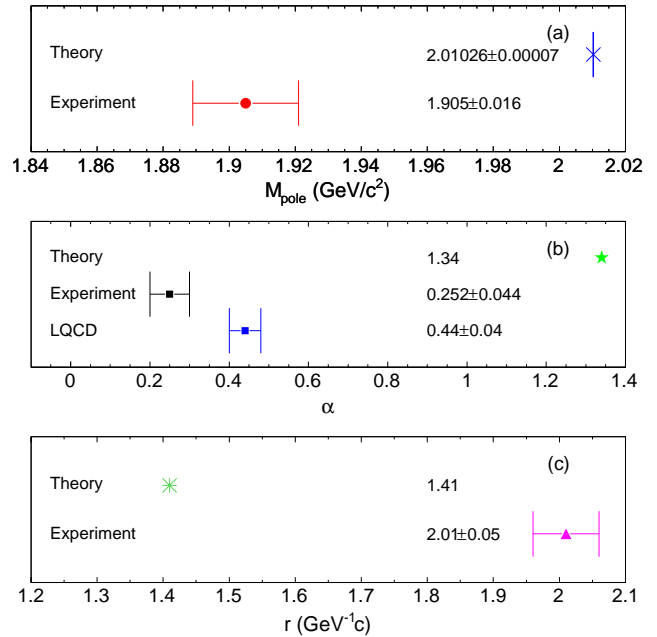}
  \caption{Comparisons of the form factor parameters determined from experimental measurements and the theoretical expectations: (a) the pole mass $M_{\rm ploe}$ in single pole model, (b) $\alpha$ in the BK model, and (c) $r$ in the ISGW2 model. }
  \label{fig:Cmp_par}
\end{figure}

When these shape parameters of the form factor parameterization are left free in the fit, the form factor parametrizations
of the single pole model, BK model, the ISGW2 model, and the series expansion model are all capable
of describing the experimental data with almost identical $\chi^2$ probability.
However, for the physical interpretation of the shape parameters in the single pole model, BK model, the ISGW2 model,
the values of the parameters obtained from the fits are largely deviated from those expected values by these models.
This indicates that the experimental data do not support the physical interpretation of the shape parameters
in these parametriziations.
Figure~\ref{fig:Cmp_par} (a), (b) and (c) show the comparisons between the measured values
and the theoretically expected values for the pole mass $M_{\rm pole}$ in single pole model,
$\alpha$ in BK model, and $r$ in ISGW2 model.
These measured parameters do not agree with the values predicted by these form factor models.

Our determined $\alpha=0.252\pm 0.044$ from this comprehensive analysis is $3.2\sigma$ smaller than
$\alpha^{\rm LQCD}=0.44\pm 0.04$
calculated in LQCD~\cite{Fermilab2005}.
Figure~\ref{fig:Cmp_par} (b) shows this comparison.

\subsection{Ratio of $f_+^{\pi}(0)/f_{D^+}$}

To check the consistency of the SM and validate the LQCD approach to the charm physics,
we can compare the ratio of the $D$ semileptonic decay form factor $f_+^{\pi}(0)$
and $D^+$ decay constant $f_{D^+}$
from both the measurements and the LQCD calculations of these quantities.
With the most accurate $f_+^{\pi}(0)_{\rm LQCD}=0.666\pm0.020\pm0.021$ calculated in LQCD~\cite{HPQCD}
and most accurate $f_{D^+}=(212.6 \pm 0.4^{+1.0}_{-1.2})$ MeV calculated in LQCD~\cite{LQCD_fD}, we get
\begin{equation}\label{eq:ratio_lqcd}
   [f_+^\pi(0)/f_{D^+}]^{\rm LQCD} = (3.13 \pm 0.14)~{\rm GeV}^{-1}.
\end{equation}
From our determined $f_+^{\pi}(0)$ given in Eq.~(\ref{eq:f0}) and $f_{D^+}$ given in Eq.~(\ref{eq:fD_exp}) (see Appendix~\ref{sec:ap}),
we find
\begin{equation}\label{eq:ratio_exp}
   [f_+^\pi(0)/f_{D^+}]^{\rm exp} = (3.11 \pm 0.08)~{\rm GeV}^{-1},
\end{equation}
which is in very good agreement within error with the LQCD prediction of the ratio give in Eq.~(\ref{eq:ratio_lqcd}).

\subsection{CKM Matrix Element $|V_{cd}|$}

Using the product $f_+^\pi(0)|V_{cd}|=0.1428\pm0.0019$ obtained from the comprehensive series expansion fit
in conjunction with the form factor $f_+^\pi(0)_{\rm LQCD}=0.666\pm 0.020\pm 0.021$~\cite{HPQCD}
calculated in LQCD for the $D\to \pi$ transition,
we extract the magnitude of the CKM matrix element $V_{cd}$ from all existing measurements of semileptonic $D$ decays to be
\begin{equation}
|V_{cd}|^{D\to \pi e^+\nu_e} = 0.2144\pm 0.0029\pm 0.0093,
\end{equation}
where the first error is from the uncertainty in experimental measurements,
the second uncertainty corresponds to the accuracy of the form factor $f^\pi_+(0)$
calculated in LQCD.
The experimental precision of this $|V_{cd}|^{D\to \pi e^+\nu_e}$ is 2 factor better than
the PDG'2014 $|V_{cd}|_{\rm PDG'2014}^{D\to \pi e^+\nu_e}=0.220\pm 0.006\pm 0.010$~\cite{pdg}
extracted from the average of the CLEO-c~\cite{CLEOc} and Belle~\cite{Belle} measurements
of $D\to\pi\ell^+\nu_\ell$ decays in conjunction with
$f_+^\pi(0)_{\rm LQCD}=0.666\pm 0.020\pm 0.021$~\cite{HPQCD}.
This big progress in improvement of the experimental accuracy
of $|V_{cd}|^{D\to \pi e^+\nu_e}$
is mainly due to the recent BESIII measurement~\cite{BESIII_D0Kenu,CKM2014_RongG_Talk},
due to BaBar measurement~\cite{BaBar}
as well as other earlier measurements~\cite{CLEO,E687,CLEOIII,CLEOII_Dp,MarkIII,BESII_D0} of
$D\to\pi\ell^+\nu_\ell$ decays.

Combining with the value
\begin{equation}
|V_{cd}|^{D^+\to\mu^+\nu_\mu}=0.2160 \pm 0.0049 \pm 0.0014
\end{equation}
extracted from both the BESIII and CLEO-c's measurements of leptonic $D^+$ decays (see Appendix~\ref{sec:ap})
we obtain the magnitude of the CKM matrix element $V_{cd}$ to be
\begin{equation}
|V_{cd}| = 0.2157\pm0.0045.
\end{equation}

\begin{figure}[!hbt]
  \includegraphics[width=0.5\textwidth]{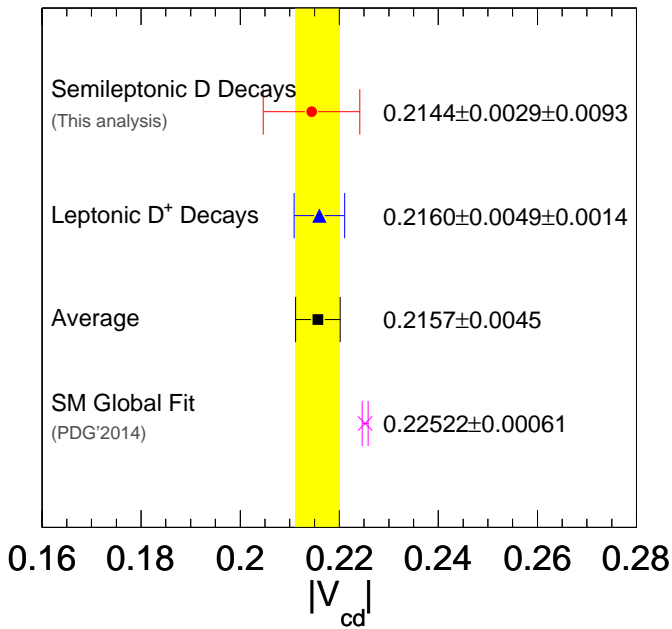}
  \caption{Comparison of $|V_{cd}|$ extracted from semileptonic $D$ decays in this analysis
with the one extracted from leptonic $D^+$ decays and along with the one from the global SM fit.}
  \label{fig:Cmp_Vcd}
\end{figure}

\begin{figure}[!hbt]
  \includegraphics[width=0.5\textwidth]{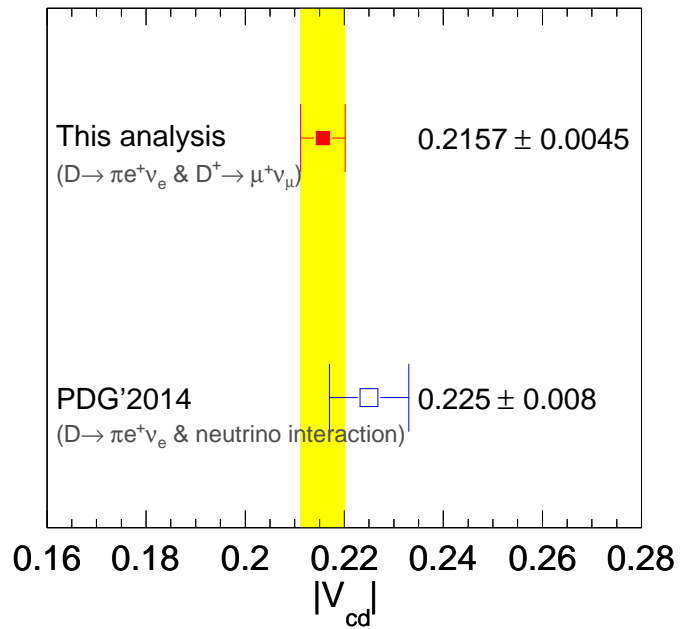}
  \caption{Comparison of $|V_{cd}|$ extracted from semileptonic $D$ decays and leptonic $D^+$ decays
           in this analysis along with the PDG'2014 value.}
  \label{fig:Cmp_Vcd_PDG}
\end{figure}

Figure~\ref{fig:Cmp_Vcd} shows a comparison of the value of $|V_{cd}|$ which is determined
with the $|V_{cd}|^{D\to \pi e^+\nu_e}$ obtained in this analysis
together with the $|V_{cd}|^{D^+\to\mu^+\nu_\mu}$ extracted from leptonic $D^+$ decays,
and the value from the global SM fit~\cite{pdg}.
Figure~\ref{fig:Cmp_Vcd_PDG} shows a comparison of our extracted $|V_{cd}|$
from all existing measurements of $D\to \pi e^+\nu_e$
and from both the BESIII and CLEO-c's measurements of $D^+\to\mu^+\nu_\mu$ decays
along with the PDG'2014 value of the $|V_{cd}|$ determined with CLEO-c and Belle's measurements
of $D\to \pi \ell^+\nu_\ell$ decays and neutrino interactions~\cite{pdg}.
Our extracted $|V_{cd}| = 0.2157\pm 0.0045$ is in good agreement within error with the PDG'2014 value
$|V_{cd}|_{\rm PDG'2014} = 0.225\pm0.008$, but improves the precision of the PDG'2014 value by over $70\%$.

\subsection{Unitarity Checks}
\begin{figure}
  \includegraphics[width=0.5\textwidth]{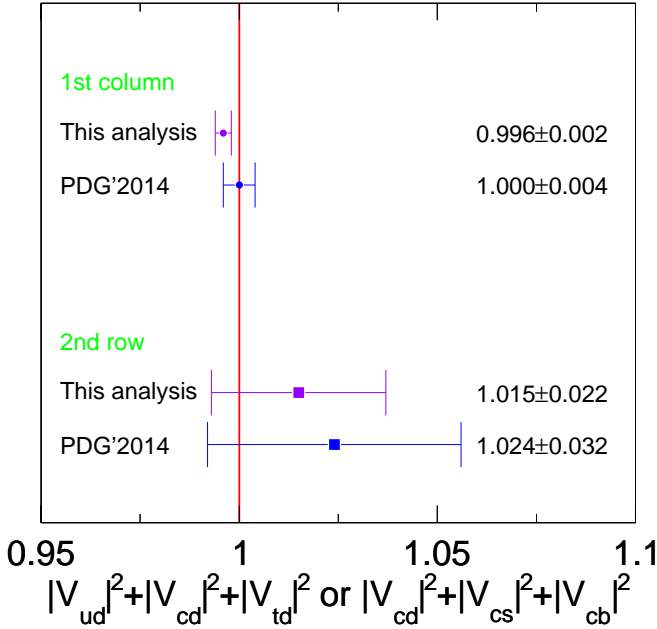}
  \caption{Unitarity checks for the first column and second row of the CKM matrix.}
  \label{fig:unitarity_check}
\end{figure}

Using the newly extracted $|V_{cd}|=0.2157\pm0.0045$,
the PDG'2014 values $|V_{ud}|=0.97425\pm0.00022$
and $|V_{td}|=(8.4\pm0.6)\times10^{-3}$~\cite{pdg},
we check the first column unitarity of the CKM matrix, which is
\begin{equation}\label{eq:U_c1}
    |V_{ud}|^2+|V_{cd}|^2+|V_{td}|^2 = 0.996\pm0.002.
\end{equation}
Using these newly extracted $|V_{cd}|=0.2157\pm0.0045$, the value $|V_{cs}|=0.983\pm0.011$
which is recently extracted from semileptonic $D$ decays and leptonic $D^+_s$ decays~\cite{Vcs_DLeptonic},
and the PDG'2014 value $|V_{cb}|=(41.1\pm1.3)\times10^{-3}$~\cite{pdg},
we find
\begin{equation}\label{eq:U_r2}
    |V_{cd}|^2+|V_{cs}|^2+|V_{cb}|^2 = 1.015\pm0.022
\end{equation}
for the second row of the CKM matrix.
The unitarity check results for the
first column and the second row of the CKM matrix
are shown in Fig.~\ref{fig:unitarity_check} together with the unitarity checks given in PDG'2014~\cite{pdg}.
The newly determined $|V_{cd}|$ and $|V_{cs}|$ give more stringent checks of the
CKM matrix unitarity compared to those in PDG'2014.

The sum of the squared matrix element in the first column of the CKM matrix deviates from the unitarity by
$$|V_{ud}|^2+|V_{cd}|^2+|V_{td}|^2 -1 = -0.004 \pm 0.002,$$
which is $2\sigma$ deviations from the unitarity.

\section{Summary}
\label{sec:sum}

By globally analyzing all existing branching fractions of the $D\to \pi e^+\nu_e$ decays
measured at earlier experiments and
products $f_+^\pi(q^2)|V_{cd}|$ measured at the Belle and the BESIII experiments
as well as the partial decay rates in $q^2$ bins measured at the BaBar and CLEO-c experiments together,
we obtain the most precise product
$$f_+^\pi(0)|V_{cd}|=0.1428\pm 0.0019.$$
From this product we determined the form factor
\begin{equation*}
    f_+^\pi(0) = 0.634\pm0.008\pm0.002,
\end{equation*}
which is in good agreement within error with LQCD calculations of the form factor,
but with more precision than the most accurate LQCD calculation of the form factor by $3.3$ factors.

We also determined the ratio of the semileptonic form factor $f_+^\pi(0)$ and $D^+$ decay constant $f_{D^+}$
to be
$$[f_+^\pi(0)/f_{D^+}]^{\rm exp} = (3.11 \pm 0.08)~{\rm GeV}^{-1}$$
from experimental measurements and
$$[f_+^\pi(0)/f_{D^+}]^{\rm LQCD} = (3.13 \pm 0.14)~{\rm GeV}^{-1}$$
from LQCD calculations.
The experimental ratio is in excellent agreement within error with that determined from LQCD calculations
of these two quantities.

Alternately, with the recently most precise semileptonic $D\to \pi e^+\nu_e$ decay form factor calculated in LQCD,
we obtain
$$|V_{cd}|^{D\to \pi e^+\nu_e}=0.2144\pm0.0029\pm0.0093,$$
where the error is still dominated by the uncertainties in LQCD calculation
of the semileptonic $D\rightarrow \pi$ form factor.
This extracted $|V_{cd}|$ is consistent within $1.1\sigma$
with $|V_{cd}|=0.22522\pm 0.00061$ from the global SM fit.

If combining this $|V_{cd}|^{D\to \pi e^+\nu_e}$
together with
$$|V_{cd}|^{D^+\to\mu^+\nu_\mu} = 0.2160\pm0.0049\pm0.0014$$
extracted from leptonic $D^+$ decays together, we find
\begin{equation*}
    |V_{cd}| = 0.2157\pm0.0045.
\end{equation*}
This newly extracted $|V_{cd}|$ improves the accuracy of the PDG'2014 determination
of $|V_{cd}|_{\rm PDG'2014}=0.225\pm 0.008$
by over $70\%$, and
is the most precisely extracted $|V_{cd}|$ from all existing measurements of semileptonic $D$ decays and
from both the BESIII and CLEO-c's measurements of leptonic $D^+$ decays up to date.

Combining the most precise $|V_{cd}|$ extracted in this work
together with other updated $|V_{ud}|$ and $|V_{td}|$ given in
PDG'2014, we find that the sum of the squared CKM matrix element in the first column deviates from unitarity
by $2\sigma$.

\section*{Acknowledgements}
This work is supported in part by the Ministry of Science of Technology of China under Contracts No. 2009CB825204; National Natural Science Foundation of China (NSFC) under Contacts No. 10935007 and No. 11305180.

\appendix
\section{Extraction of $|V_{cd}|$ from Leptonic $D^+$ Decays}
\label{sec:ap}

In this appendix, we present the determination of $|V_{cd}|$ by analyzing the existing measurements of leptonic $D^+\to\mu^+\nu_\mu$ decays.

In SM of particle physics, the branching fraction for $D^+\to\mu^+\nu_\mu$ decay is given by
\begin{eqnarray}\label{eq:BF_lep}
  B(D^+\to\mu^+\nu_\mu) & =  & \frac{G_F^2}{8\pi} \tau_{D^+} m_\mu^2 m_{D^+} \left(1 - \frac{m_\mu^2}{m_{D^+}^2} \right)^2
  \nonumber
  \\
  && \times f_{D^+}^2 |V_{cd}|^2,
\end{eqnarray}
where $\tau_{D^+}$ is the lifetime of $D^+$ meson,
$m_\mu$ is the mass of muon and $m_{D^+}$ is the mass of $D^+$ meson.
The parameter $f_{D^+}$ is the decay constant, which is associated with the strong interaction effects between the two initial-state quarks.

In 2008, the CLEO-c Collaboration accumulated $460055\pm 787$ $D^-$ tags
by analyzing 818 pb$^{-1}$ data taken at 3.773 GeV and selecting $D^-$ mesons from
6 hadronic decay modes
of the $D^-$ meson.
They observed $149.7 \pm 12.0$ signal events
for $D^+ \rightarrow \mu^+\nu_\mu$ decays in the system recoiling against these $D^-$ tags,
and measured the branching fraction
$B(D^+ \rightarrow \mu^+\nu_\mu)=(3.82 \pm 0.32 \pm 0.09)\times 10^{-4}$
~\cite{cleo-c_fD_2008}.

In 2014, the BESIII Collaboration investigated the
$D^+ \rightarrow \mu^+\nu_\mu$ decays by analyzing 2.92 fb$^{-1}$ data taken at 3.773 GeV.
From 9 hadronic decay modes of $D^-$ meson, the BESIII Collaboration accumulated $1703054\pm3405$ $D^-$ tags.
In this $D^-$ tag sample they observed $409.0\pm 21.2$ signal events for
$D^+ \rightarrow \mu^+\nu_\mu$ decays and measured the branching fraction
$B(D^+ \rightarrow \mu^+\nu_\mu)=(3.71 \pm 0.19 \pm 0.06)\times 10^{-4}$~\cite{BESIII_Dptomunu}.

Averaging these two branching fractions, we obtain
\begin{equation}\label{eq:bf_Dptomunu}
B(D^+\to\mu^+\nu_\mu) = (3.74\pm0.17)\times10^{-4},
\end{equation}
where the error is the combined statistical and systematic errors together.

Inserting the values
$m_\mu=(105.6583715\pm0.0000035)$ MeV,
$m_{D^+}=(1869.61\pm0.10)$ MeV,
and $\tau_{D^+}=(1040\pm7)\times10^{-15}$ s,
from PDG'2014~\cite{pdg} and the average value of branching fraction given in Eq.~(\ref{eq:bf_Dptomunu})
into Eq.~(\ref{eq:BF_lep}),
the product of the decay constant and the magnitude of CKM matrix element $V_{cd}$ is determined to be
\begin{equation}\label{eq:fDVcd}
    f_{D^+}|V_{cd}|=(45.92 \pm 1.04 \pm 0.15)~\rm MeV,
\end{equation}
where the first error is from the statistical and systematic uncertainties in the measured branching fractions,
and the second error is due to the uncertainties in the masses of muon and $D^+$ meson, the lifetime of $D^+$ meson.

Dividing the product $f_{D^+}|V_{cd}|$ by the value $f_{D^+}=(212.6\pm0.4^{+1.0}_{-1.2})$ MeV
which is the newest and most precise value of decay constant calculated in LQCD with $N_f=2+1+1$ quark flavors~\cite{LQCD_fD},
we obtain
\begin{equation}
    |V_{cd}|^{D^+\to\mu^+\nu_\mu}=0.2160\pm0.0049\pm0.0014,
\end{equation}
where the first error is from the statistical and systematic uncertainties in the measured branching fractions,
and the second error is mainly due to the uncertainties in the lifetime of $D^+$ meson, and the $f_{D^+}$ calculated in lattice QCD.

Alternatively, by inserting $|V_{cd}|=0.22522\pm0.00061$ from the global SM fit~\cite{pdg} into Eq.~(\ref{eq:fDVcd}),
we determine
\begin{equation}\label{eq:fD_exp}
    f_{D^+}=(203.9\pm4.6\pm0.9)~\rm MeV,
\end{equation}
which is the most precisely determined $D^+$ decay constant based on the branching fractions
for $D^+ \rightarrow \mu^+\nu_{\mu}$ decays measured at both the BESIII and CLEO-c experiments.

\end{document}